# High-resolution TEM imaging on 3D nanocrystals with tilt illuminations

Tsumoru Shintake[1]*

*[1]OIST: Okinawa Institute of Science and Technology Graduate University*

*1919-1, Tancha, Onna-son, Okinawa 904-0495 Japan*

*E-mail: shintake@oist.jp

When we observe 3D objects, such as nanocrystals using a transmission electron microscope (TEM), the significant depth of the sample causes the contrast transfer function (CTF) to mix, resulting in complex and confused image. This paper predicts that a high-resolution projected image can be obtained by tilting the illumination and filtering it with a ring slit. This method produces images similar to those obtained using scanning transmission electron microscopy (STEM), which is not affected by the CTF problem. Potential applications are such as MOFs (metal-organic frameworks), perovskites and Solid-State Electrolyte (SSE) for Li-ion battery, where small ions or molecules play important role.

## 1. Introduction

MOFs (metal-organic frameworks), perovskites and Solid-State Electrolyte (SSE) are solid-state materials and systems that offer superior safety and thermal stability compared to conventional components[1-3]. They have tunable crystalline structures and fast ionic/electronic transport, and thus essential for high-performance energy storage and conversion systems, such as batteries, solar cells, fuel cells and capacitors.

Due to its dimensional flexibility and tunability, the direct observation method based on a transmission electron microscope (TEM) may be useful tool to study structure same as conventional techniques such as X-ray crystallography or nuclear magnetic resonance (NMR). However, when we observe such nanocrystals using a transmission electron microscope (TEM), the sample causes the contrast transfer function (CTF) to mix, resulting in complex and confused image[4]. In order to solve the problem, the author proposes a new microscopy as shown in Fig. 1. The conventional microscopy shown in Fig. 1(a), is basically point-to-point projection system, where the lens captures spherical waves emitted from the point on the object, followed by focusing down to point on the image. TEM is in this category. The new method (b) is based on plane wave imaging. As our sample is a 3D crystal, it effectively reflects the plane wave due to Bragg diffraction, acting as a reflective mirror. The reflection angle is chosen by ring-slit and focused down through the lens onto the image. Note that the incoming electron beam, i.e. the illumination, also passes through the ring-slit and provides a reference wave for the diffracted wave, causing an interference effect around the image point. As will be discussed later, wave interference creates a 3D volume hologram of the image elongated along the z-axis (the optical axis), which is why we have a projected image. This is a new type of telecentric microscopy[6)] and may therefore be suitable for tomography imaging for thick specimens. In TEM, the CTF function is caused by the interference effect of spherical waves, similar to the effect seen in Newton rings. In new microscopy, however, there are no spherical waves, so the CTF is not observed in principle.

Figure 2 illustrates the practical implementation of the new imaging method using TEM. The hardware configuration is essentially the same as that of a conventional TEM. In this case, however, the electron beam is steered to create off-axis illumination of the sample. This is followed by a ring slit to select the desired Bragg diffraction for projection imaging. It also cut unwanted noise signal to improve image quality. In reality, the sample is located in the middle of the magnetic pole piece of the objective lens. The ring-slit is also located within the pole piece, right after the sample. The physics of diffraction and image formation are described in more detail in subsequent sections.

## 2. Plane wave scattering by a crystal structure

In contrast to the way an object reflects light under white-light ambient illumination, a crystal by X-rays or coherent electron beams only causes diffraction when the selection rule, i.e. the Bragg condition, is satisfied. The condition is

$$d = \frac{\lambda}{2\,sin\theta} \qquad (1)$$

The wavelength is represented by $\lambda$, the reflection angle by $\theta$, and the spacing of the periodic structure by $d$. By reversing eq. (1), we find matching condition between the reciprocal lattice vector: $1/\underline{d}$ and the wave number: k, which is graphically illustrated in Fig. 3.

The incident wave vector $\mathbf{k}_1$ is scatted into $\mathbf{k}_2$ by the reciprocal lattice point ($h$,$k$,$l$) of the crystal. The scattering vector is given by

$$\mathbf{S} = \mathbf{k}_2 - \mathbf{k}_1 \qquad (2a)$$

$$|S| = 2k_0\, sin\theta \cdot sin\frac{\alpha}{2} \qquad (2b)$$

We assumed there is no energy gain or loss in the scattering process, and thus $\mathbf{k}_1$ and $\mathbf{k}_2$ have the same vector length $k_0$. Even if the scattering angle changes, $\mathbf{k}_2$ keeps the same length, and thus the end point will be on a sphere called Ewald's sphere[7)].

magine placing a spherical object, such as a basketball, in tap water. The boundary between the water's surface and the ball forms a circle. If we select only the diffraction wave vector on this circle using the ring slit, the scattering vector will be in the xy-plane (i.e. the z-component will be zero). Since the scattering vector represents the interaction between the electron beam and the lattice vector within the crystal, the lattice is chosen to be parallel to the beam axis. And thus, it creates a longitudinal projection image by means of the volume hologram.

Equation (2b) shows how the length of a vector varies due to the azimuthal angle of $\alpha$. Near to the illumination spot, i.e., $\alpha \sim 0$, the scattering vector becomes small, and thus the lattice spacing in real space becomes very wide. At opposite $\alpha = \pi$,, the scattering vector becomes minimum:

$$|S| = 2k_0\, sin\theta \qquad (3)$$

This is commonly known as Bragg condition, which is identical to Eq. (1).

## 3. Interpretation of laser reflection from a needle in the context of Bragg diffraction

In this section, we will use an analogous demonstration with more practical tools — a laser and a thumbtack pin — to explain our imaging method.

In Fig. 4, the author shone a laser pointer on a pin at an oblique angle from behind. The green circle appeared on a white wall, which was roughly one meter away from me. I carefully positioned the pin so that it was perpendicular to the wall, i.e., the pin was heading to the center of the circle. The diameter of the green circle is determined by the distance to the wall and the angle at which the pin is shined on.

The laser beam is larger than the pin, so most of the laser power passes the pin and shines on the left-hand side of the green circle. This makes it look like a diamond ring during a solar eclipse. It corresponds to the illumination, or the reference wave vector: k0 in Fig. 3.

You might be wondering about the origin of the circle. The simple answer is that the cylindrical mirror creates a ring-shaped reflection. Refer to Fig. 5. The thumbtack pin is a cylindrical mirror which is magnified and represented by a series of narrow rectangular mirror segments. As these mirrors have different azimuthal angles, the reflected beam goes in different directions. Please note that the reflected beam rotates twice as fast as the mirror. This results in a ring-shaped beam of light being reflected from half of the cylinder.

Now we discuss how to connect with Bragg diffraction phenomena, with refer to Fig. 5. Assume the mirror segment is a multi-layer mirror, whose layer spacing is tuned by Bragg condition. As we learned from Eq. (1), small-angle reflection is generated by wide-spaced multilayers, while large-angle reflection is due to small spacing. The minimum spacing corresponds to 180-degree back reflection.

This is a thought experiment, where we create a crystal structure by moving and overlapping the multi-layer structure. The reflection angle is still governed by the Bragg condition, so the reflection and the circle on the wall will not change. There are a few more important points to note. As shown in the side view (bottom left) of Fig. 5, all mirror segments are parallel to the pin, and the scattering vector S lies in the xy plane. This is the same situation as the Bragg diffraction passing through the ring slit shown in Fig. 3. Therefore, we can conclude that the diffraction passing through the ring-slit originates from the longitudinal lattice planes in the crystal sample. And thus, the volume hologram provide projection image of the sample.

## 4. Conclusions

In this theoretical study, we discussed a new type of microscopy.

(1) TEM with a ring slit positioned behind the sample collects Bragg diffraction and forms a volume hologram at the detector.

(2) Analogous demonstration was introduced to explain basic concept of this new imaging method.

(3) It acts as a telecentric microscope, making it suitable for imaging nanocrystals. It may contribute to superior imaging with a tomography microscope.

(4) Possible applications include metal-organic frameworks (MOFs), perovskites and solid-state electrolytes (SSEs). It can also be used to image small protein crystals.

## Figures

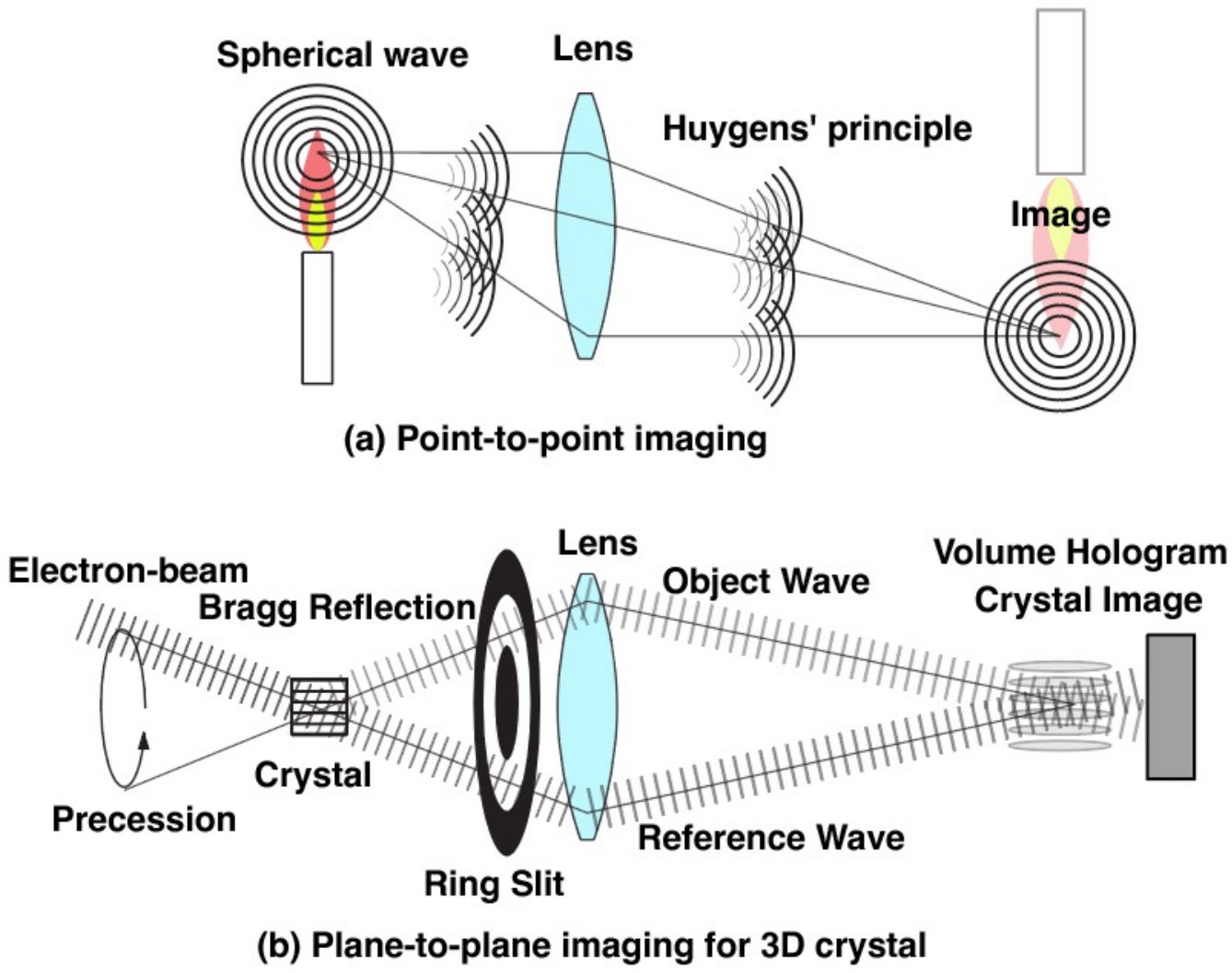

**Fig. 1.** Basic concept: (a) Common point-to-point imaging. (b) The new plane-to-plane imaging concept for 3D crystals. The created volume hologram represents the projected image of the crystal sample.

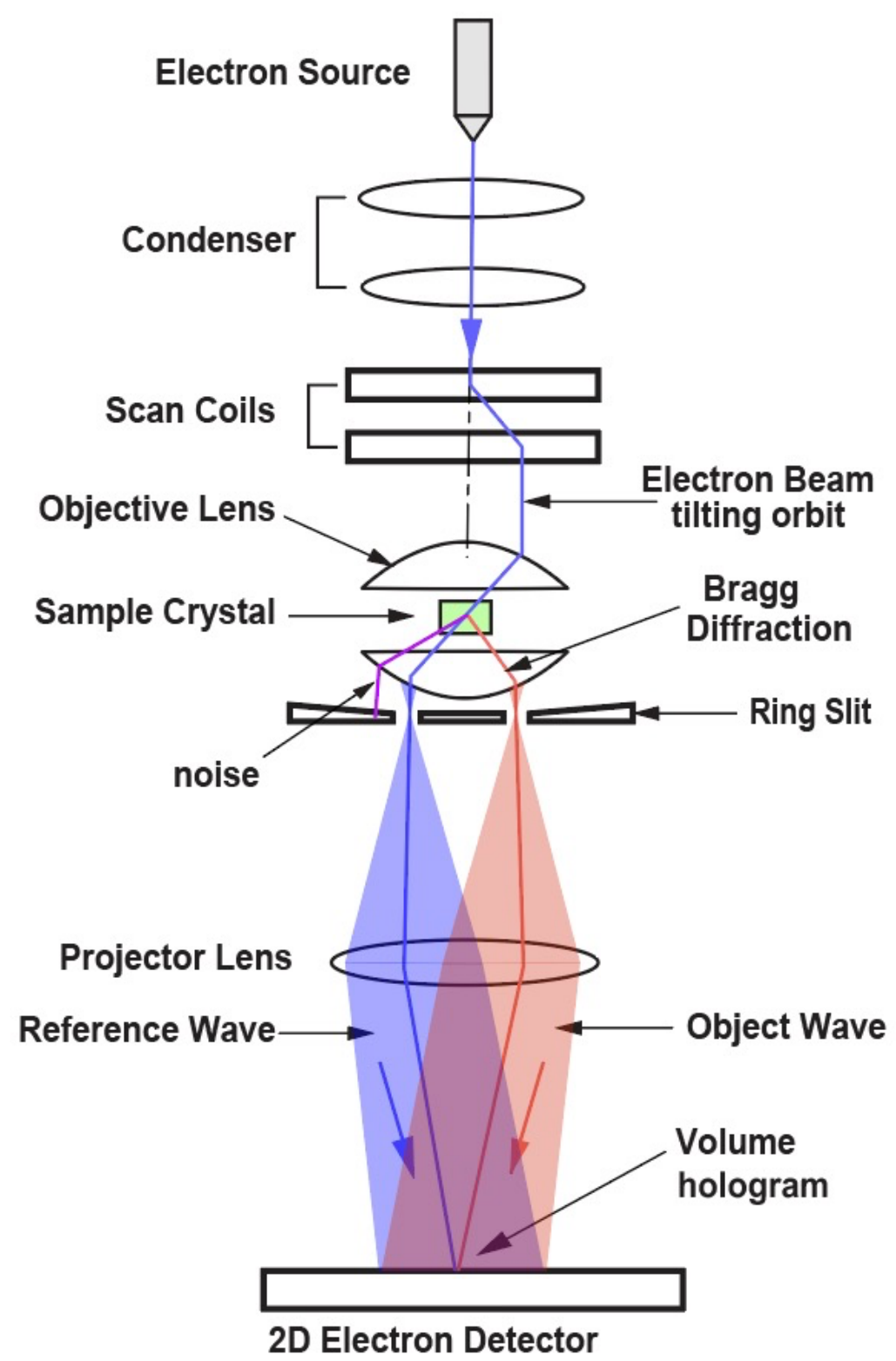

**Fig. 2.** The electron microscope with tilt illumination for the crystalline sample.

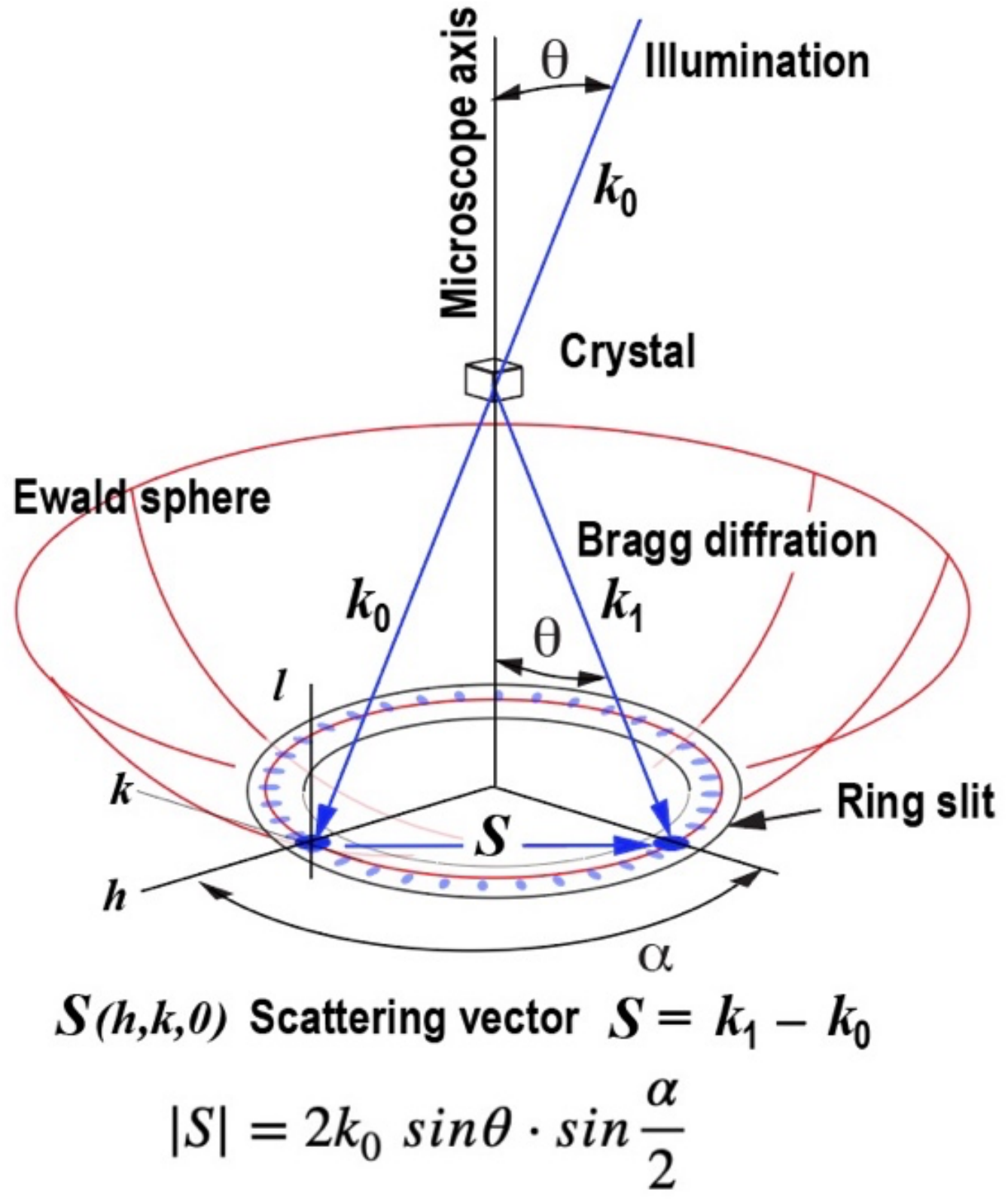

**Fig. 3.** Wave vector relation between incoming illumination and Bragg diffraction. The scattering vector gives the structure information inside the crystal sample.

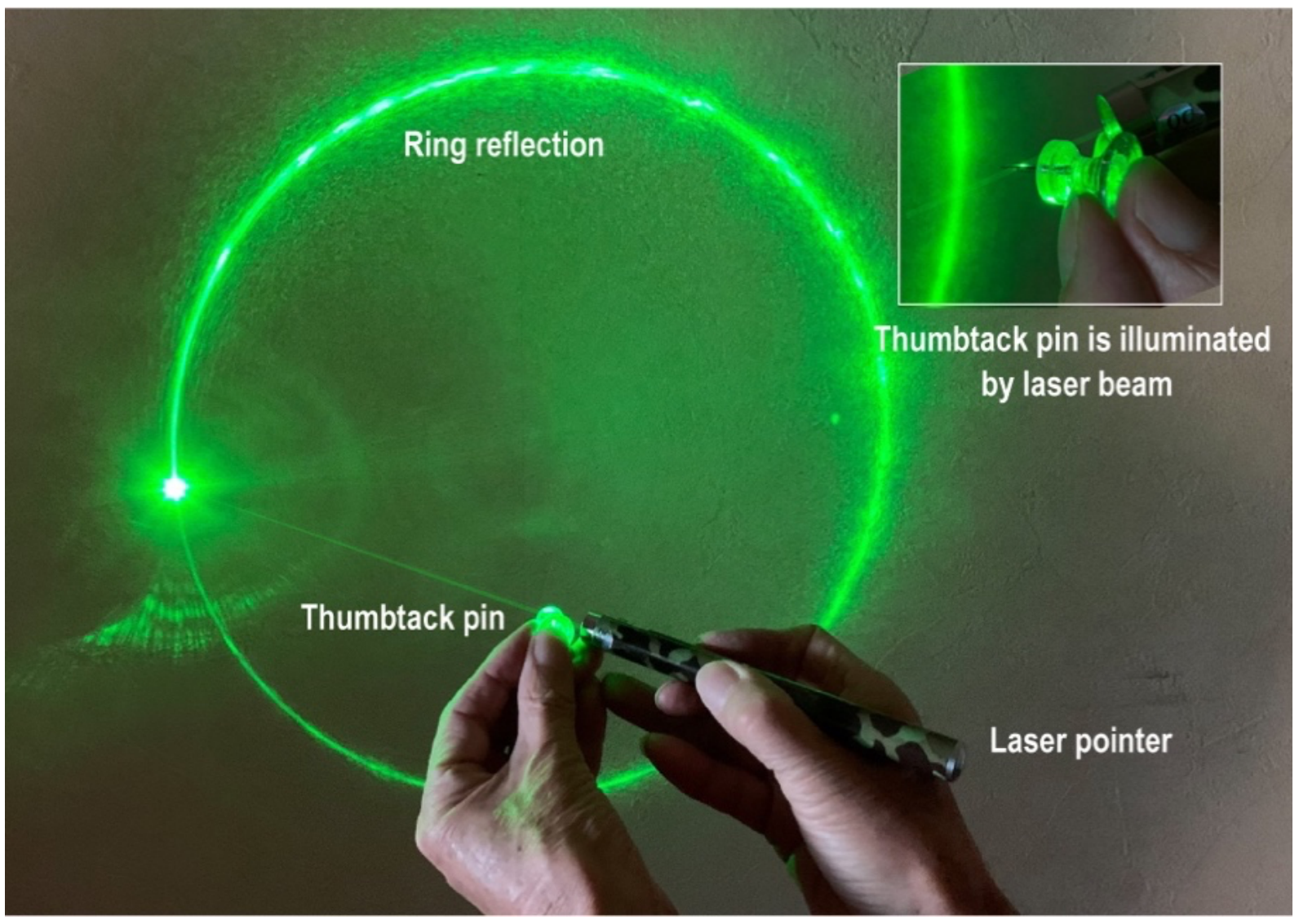

**Fig. 4.** Analogous demonstration of wave scattering from a small needle.

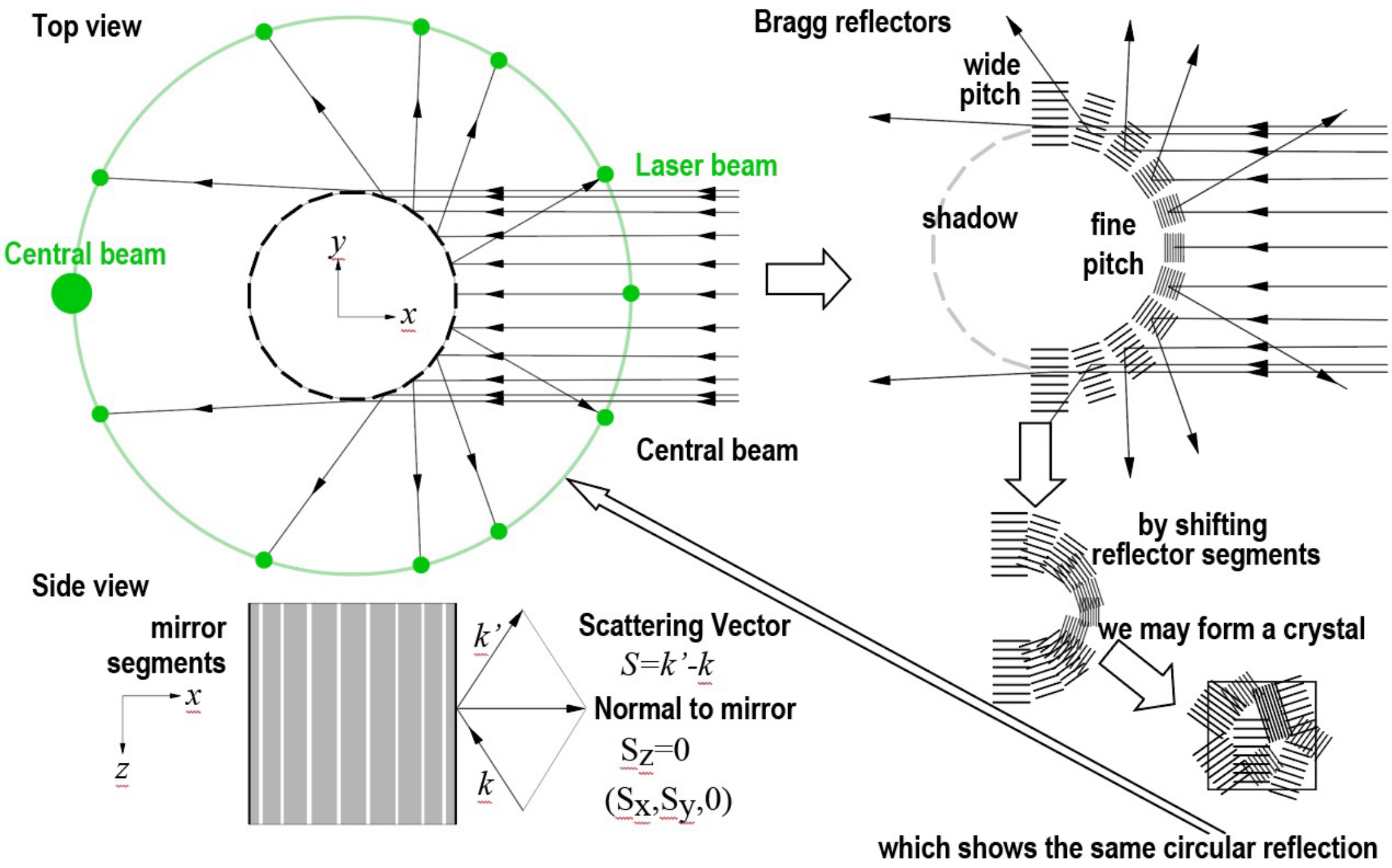

**Fig. 5.** Interpretation of laser reflection from a needle in the context of Bragg diffraction.